# Evaluating the SharedCanvas Manuscript Data Model in CATCHPlus

## Robert Sanderson[1], Hennie Brugman[2], Benjamin Albritton[3], Herbert Van de Sompel[1]

1. Los Alamos National Laboratory
Los Alamos, NM 87544
USA
{rsanderson,herbertv}@lanl.gov

2. Meertens Institute
Joan Muyskenweg 25
1096 CT Amsterdam
The Netherlands
hennie.brugman@meertens.knaw.nl

3. Stanford University
Stanford, CA 94305
USA
blalbrit@stanford.edu

**Abstract**

In this paper, we present the SharedCanvas model for describing the layout of culturally important, hand-written objects such as medieval manuscripts, which is intended to be used as a common input format to presentation interfaces. The model is evaluated using two collections from CATCHPlus not consulted during the design phase, each with their own complex requirements, in order to determine if further development is required or if the model is ready for general usage. The model is applied to the new collections, revealing several new areas of concern for user interface production and discovery of the constituent resources. However, the fundamental information modelling aspects of SharedCanvas and the underlying Open Annotation Collaboration ontology are demonstrated to be sufficient to cover the challenging new requirements. The distributed, Linked Open Data approach is validated as an important methodology to seamlessly allow simultaneous interaction with multiple repositories, and at the same time to facilitate both scholarly commentary and crowd-sourcing of the production of transcriptions.

## Introduction

There are many digital repositories that maintain hundreds of thousands of page images of medieval manuscripts or other historically important, handwritten documents. These images act as surrogates for physical objects and are often the only way in which scholars and students can interact with the material, making it essential that the experience be rich with access to as much of the research about the physical objects as possible.

Repositories have two main tasks in this domain, digitization of the original and then providing access to the digital resources. While the former is a process specific to the institution, it was recognized that the latter would benefit greatly from a shared data model and implementations to facilitate the creation of software environments to consume that model. It was also recognized that the knowledge about the object is often maintained across multiple repositories, and any sustainable solution would need to embrace this aspect rather than continue with current "silo" based approaches, where a single repository holds all of the information.

In this paper, we describe the SharedCanvas data model for handling digitized manuscripts that covers all of the use cases known during the design process. We then introduce two further collections from CATCHPlus with their own requirements, and attempt to use the model to fulfill them as a method of evaluation.

## SharedCanvas Data Model

### Requirements

The SharedCanvas data model (Sanderson 2011) was developed in this context to provide a distributed solution to the real world use cases of several of the largest repositories of digitized medieval manuscripts. These use cases outline the challenges of the multiple relationships between the digital resources and the physical objects they are derived from, with scholarly commentary as a further layer to consider. The resources involved comprise:

1. Multiple texts, such as transcriptions, editions and translations of the object's intellectual content.
2. Multiple images of the object, per page, at different resolutions or in some cases using multi-spectral imaging techniques.
3. Transcriptions of non-text, such as musical scores transcribed into XML, equations transcribed into LaTeX, or diagrams into more comprehensible images.
4. Audio files, that can be considered realizations of the above musical scores, or potentially video files as realizations of musical or theatrical descriptions.
5. Scholarship and automated analysis concerning all of the above.

While it might seem that the most common case for manuscripts is to have an image of the full page, according to an estimate of Christopher de Hamel, the Donnelley Fellow Librarian at Corpus Christi College, Cambridge, less than 1% of existing medieval material has been digitized. Further, within this 1%, there are many cases in which only fragments of a manuscript remain to be imaged, or when digitization of a page is impossible because of the damage it would cause. However, although only part, or none, of the page might be digitized, the text may still be known from other witnesses, earlier transcriptions or non-destructive physical inspection. Thus, the oft-constructed relationship between image and text is overly simplistic, as both are aspects of a digital surrogate for the physical object.

The association of entire resources with each other also only covers the most basic of use cases. It must be possible to associate part of a complete text with the page from which it was transcribed. Equally, only part of an image may depict a page with the rest being either the other half of an open spread of a manuscript, or just the scanning bed, ruler and color charts which are often not of interest.

Furthermore, the sequence of the folios and quires often changes over time due to rebinding of the manuscript, pages being cut out, or the disbinding and subsequent

dispersal of all of the individual leaves. This presents further challenges:
1. It must be possible to describe the different orderings without duplicating all of the information needed for presentation.
2. The descriptions must allow the seamless inclusion of resources from different repositories, as the leaves may be physically and institutionally separated.
3. It must be possible to discuss and digitally annotate the manuscript in context of specific orderings.

**Data Model and Architecture**

By adopting the principles of Linked Open Data (LOD) and the Architecture of the World Wide Web (WWW), the SharedCanvas model recognizes the importance of distributed information creation, management and ownership. A single description of all available constituent resources of a digital surrogate for the physical object can be constructed using LOD where, for example, the text transcriptions are derived from one repository and the digital images come from another. This ability to distribute the creation and maintenance of the resources used to build the digital surrogate is of key importance in many of the requirements, but does not in itself provide a solution for multiple alignments of images, texts and other resources.

The solution proposed for the required complex inter-relationships uses a Canvas paradigm, such as in standards including HTML5, SVG and PDF. The Canvas is an abstract space that stands for a single page, and has dimensions relative to the physical object: the top left hand corner of the Canvas represents the top left hand corner of a rectangular bounding box around the page, and likewise for the bottom right hand corner. By giving the Canvas a globally unique identifier (an HTTP URI following the Architecture of the WWW), we are able to make statements about it without needing to have access to a single descriptive metadata file. Thus, in the same way that anybody can create a hypertext link from one web page to another, anybody can associate additional information with the Canvas.

Annotations are used as the means of associating additional information with the Canvas. This information could be scholarly commentary in the intuitive sense of an annotation, however the same approach allows the depictions, transcriptions, realizations and other resources needed to build up a rich digital surrogate to be linked. The requirements of the domain, such as associating part of an image with part of the Canvas, must then be available within the Annotation model used.

**Open Annotation Data Model**

As will become apparent during the evaluation section of this paper, the choice of Annotation model is of the utmost importance. If the underlying Annotation model lacks certain features, then the SharedCanvas model that builds on it will have a much more difficult time in fulfilling requirements. As such, a description of the Open Annotation data model, used as the basis for SharedCanvas, is warranted.

The Open Annotation data model has three primary resources:

1. Annotation: A document that describes the association of a Body and one or more Targets. [Node labeled 'Anno' in Figure 1]
2. Body: The comment or source of information that is somehow "about" the Target(s). [Node labeled 'Body' in Figure 1]
3. Target: The resource that the Body is concerned with. [Node labeled 'Tgt' in Figure 1]

Unlike in previous systems, the Body may be of any media type; a video comment is treated in the same way as a textual comment at the modeling level.

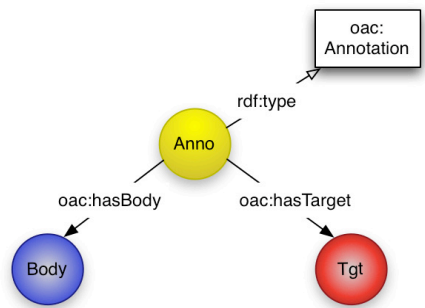

Figure 1. Basic Open Annotation data model

Open Annotation also allows for parts of resources to be used as either the Body or Target in an Annotation. For example, part of a video could be associated with the region of an image that is being discussed at that point. This is achieved with the inclusion of three additional types of resource:

1. Constraint: A resource that <u>describes</u> the segment of the main resource (either Body or Target) that is of interest, such as an SVG path for an image, or a character offset and length for plain text. [Nodes labeled 'Con1' and 'Con2' in Figure 2]
2. ConstrainedTarget: A resource that <u>identifies</u> the segment of interest within a Target, such as the region of an image being discussed. [Node labeled 'ConTgt' in Figure 2]
3. ConstrainedBody: A resource that <u>identifies</u> the segment of interest within a Body such as the part of the video in which the image is discussed. [Node labeled 'ConBody' in Figure 2]

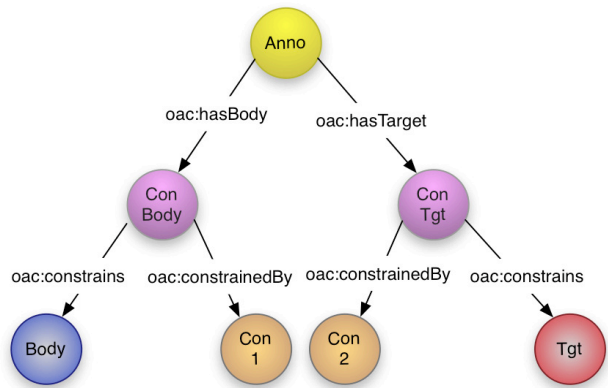

Figure 2. Complete Open Annotation Data Model

Open Annotation follows the LOD guidelines, and thus the Annotation itself is a resource on the web that can be discovered, retrieved, indexed and searched in the same way as any other online resource. It can equally be protected in the same way, so that only authenticated users can access it, if this is a requirement. This also means that the Body is a separate resource from the Annotation with its own URI.

**SharedCanvas's Use of Open Annotation**

The Canvases are built up collaboratively via the distributed creation of Annotations that associate images, texts and other resources from anywhere on the Web with the Canvas. The body of the Annotation is the transcription, image, audio or commentary and the target of the annotation is the Canvas, or that part of the Canvas where the resource should be rendered. This is depicted in Figure 3, where the blue node represents the Annotation associating the image with the full canvas, and the purple Annotation node at the top associates the transcription with the part of the Canvas that represents the area of the page where the text appears. These Annotation nodes are the specializations of the oac:Annotation node, depicted in yellow with the label 'Anno' in Figures 1 and 2. Further types of Annotation can be created for more specialized use cases, such as to distinguish between the transcription of text and the transcription of musical notation.

associated rdf:List construction. This same approach can be used to identify sub-sections of the sequence, called Ranges (Rng in Figure 4), in order to describe chapters, texts, quires or any other hierarchy of interest.

OAI-ORE Aggregations are further applied to aid in discovery of the Annotations that associate resources with the Canvases, and a top-level Manifest collection to pull everything together. This discovery layer of resources is only one of many possible configurations, as different use cases may require or benefit from alternative distributions of the Annotations. The default division is by the type of resource being annotated onto the Canvas, generating groupings such as a TextAnnotationList and an ImageAnnotationList. Another option might be to group all Annotations, regardless of type, together per folio or quire. There might also be additional nested collections, such that all of the texts are grouped together by folio, and then all of those groups are collected into a single ordering for the folios.

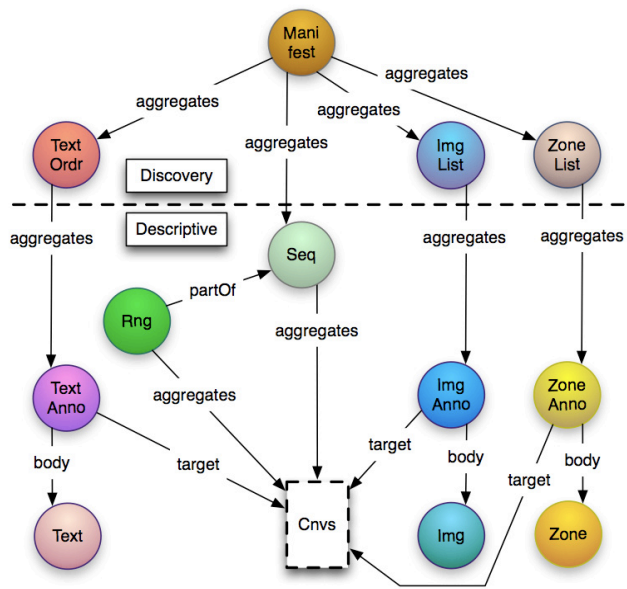

Figure 4. SharedCanvas Data Model

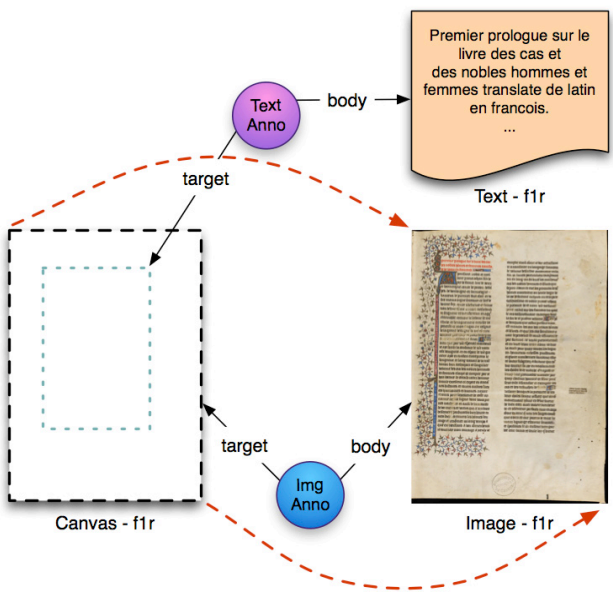

Figure 3. Annotating a Canvas with Resources

The data model also includes the notion of a Zone, which has dimensions in the same way as a Canvas, but lacks the semantic association with any one physical page. A Zone can be annotated with resources, and then in turn annotated on to one or more Canvases, perhaps with additional manipulation such as rotation or scaling. The Zone is thus able to maintain a group of Annotations along with their relative layout. This means that the information does not need to be repeated across Canvases that share some of the same resources.

**SharedCanvas's Use of OAI-ORE Aggregations**

The multiple orderings requirement is handled by collecting the identifiers for the Canvases into different Sequences, each describing one of the orders. A Sequence ('Seq' in Figure 4) is a specialization of an OAI-ORE Aggregation (Van de Sompel 2009) which maintains the order of its aggregated resources through the use of an

## Target Collections

Two collections are used in the current research to evaluate the SharedCanvas data model. These collections were not used in the original design, and hence their use cases may provide further requirements. The extent to which the model can support these collections is a strong indicator as to whether it is sufficiently robust for general use.

The context of this research is the CATCHPlus project. The CATCH scientific research program develops generic methods and techniques cutting across the areas of the humanities and computer science, aiming to help unlock the digital collections of cultural heritage institutions. CATCHPlus was started in 2009 to turn research prototypes and demonstrators from CATCH into a coherent set of production quality tools and services for the cultural heritage sector.

**Archive of the Queen's Office**

The CATCHPlus subproject Scratch4All deals with searching in scanned handwritten documents, on basis of the combination of machine learning and the partial manual transcription at line or word level. The primary collection for Scratch4All consists of index books of the Queen's cabinet[i] ("Kabinet der Koningin", or KdK) that are digitized and made available by the Dutch National Archive. The Archive of the Queen's Office itself contains the official documents of the Dutch State and are therefore of great historical value.

Currently, the collection consists of 40.000 high-resolution TIFF images and transcriptions with either automatically generated bounding boxes for the lines of text, or manually generated zones for individual words, providing an ideal test case for SharedCanvas. Exploring the inter-relationship of textual structure, such as paragraphs and sentences, and physical structure, such as words and lines, is of importance in this regard.

In a second, experimental, use case for this collection, the transcription texts themselves will be automatically annotated with the help of a Named Entity Recognition service. To this end, existing NER software for Historic Dutch will be used to add Open Annotations to segments of the texts in the SharedCanvas representations of the documents.

A final modeling issue to be investigated is that of templates for document layout. The KdK documents are semi-structured, with each page having a standard layout. It may be possible and useful to represent this standard layout once for multiple pages, rather than to repeat it for every page.

**Sailing Letters**

The Sailing Letters[ii] is a collection of 38.000 Dutch letters and documents from the 17$^{th}$ and 18$^{th}$ century, which were obtained by English pirates, and are now archived by the UK National Archives. A number of scientific and cultural heritage institutions in The Netherlands are collaborating on digitization efforts and exploitation of this material. Access will be free for scholarly use.

The use of the SharedCanvas data model will consist of two synchronized levels of transcription (a literal 'diplomatic' transcription and a normalized 'critical' edition) associated with the images via the Canvas. A user interface requirement is that the appropriate segments of the two texts and of the image should all be highlighted when a user selects a section from any one of the three.

Additionally, names will be annotated separately. The collection will then be used to perform historic, literary and linguistic research on the document images, metadata and annotations.

The Sailing Letters collection poses several interesting challenges to the SharedCanvas model:

1. Bounding boxes for lines in the scanned images will be automatically detected by a 'line strip detection' web service. This will require preprocessing at the image level (cropping, rotation, selection of 'blocks of text' and several more advanced image transformations). This means that different co-ordinate systems might be used if annotations are directly attached to different images.
2. Scanned images sometimes contain multiple pages, or pages with complex layout, as in Figure 5. Such layouts require rotation of segments for reading ease, overlapping and non-rectangular bounding boxes for text, and could benefit from interactive digital folding and unfolding of the letter.
3. Visualization of Sailing Letters scans with associated 'layers' of transcription texts is challenging. Explicit modeling of the text at several document levels (pages, text blocks, lines, words, etc) may help generate more readable text representations.

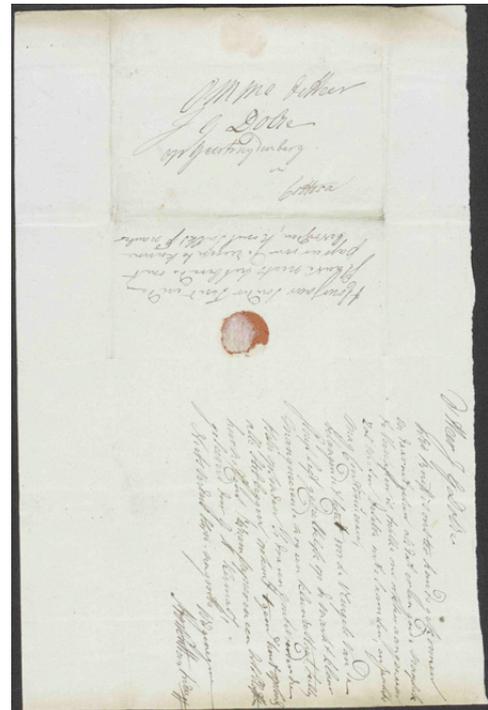

Figure 5. Example from Sailing Letters Collection

## Evaluation

The evaluation of the use cases against the data model is divided into two sections: the requirements that can be fulfilled solely with SharedCanvas constructions, and those which are either out of scope for a layout oriented model, or can more easily be solved with constructions from the underlying Open Annotation model.

**SharedCanvas Specific Requirements**

The basic use cases put forward by these collections, such as alignment of transcription and images, are all covered by SharedCanvas, as described in the Requirements section of the Data Model description, or in previous work. The new use cases are investigated and the solutions described in detail in order to evaluate the suitability of the model for further unknown collections. These requirements are:

1. Multiple Coordinate Frameworks
2. Alignment of Multiple Texts
3. Interaction with Canvas Segments

**Multiple Coordinate Frameworks**

The first challenge is the existence of multiple coordinate frameworks for the same Canvas, due to image manipulation and analysis. As a line detection service may need to rotate the image in order to accurately discover the line segments, these segments will not map directly back to the unrotated image annotated onto the Canvas.

There are two possible solutions, one theoretical and the other more practical. In theory, the Constraint model of Open Annotation can include any information desired concerning the manipulation of the resources, either for transformation or segmentation. In this way, all of the operations performed by the line detection service upon the base image could be recorded in Constraints. However, only the most sophisticated of consuming environments would be able to recreate these operations, and thus this is not a practical solution.

As there is a single Canvas with dimensions that represent the physical page, any segments of that Canvas must be expressed in its own coordinate framework. The image transformations of the line detection system could be reversed before the segment descriptions are published, so that the transformed image is only ever used internally. Reversing the rectangular regions from a transformed space results in polygons in the original set of dimensions, and the line detection system could simply publish these polygons. This demonstrates how SharedCanvas constructions can be used for an output format of an interoperable service, creating the input for the viewing environment.

In SharedCanvas, polygonal segments are supported via the SVG standard, which would be exposed as an SVGConstraint, and in this case as part of a TextAnnotation. All of the major web browsers support SVG natively. This solution maintains the scope of SharedCanvas as a layout oriented system, consumable in common environments, rather than requiring complex image manipulation capabilities. This is demonstrated in Figure 6, where the bounding box service above the dashed line has to rotate the image by 3 degrees clockwise to create a rectangular box, but can export that rectangle as a polygon in the original image's coordinate framework, which is equivalent to the framework of the Canvas. A TextAnnotation can then use the SVGConstraint to align its text with the appropriate segment of the Canvas.

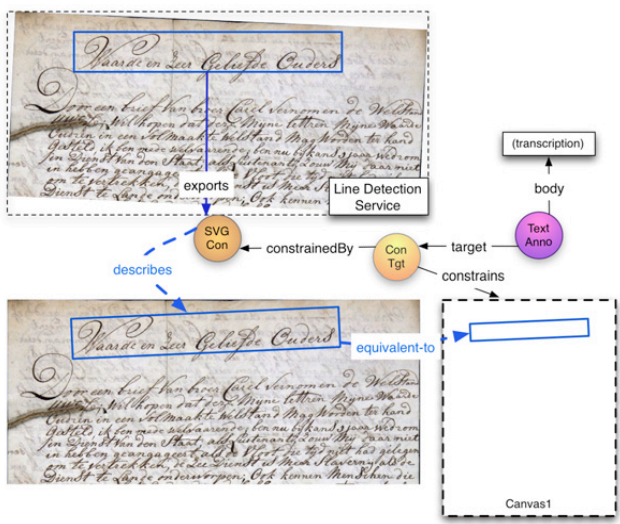

Figure 6. Polygonal bounding areas for text

### Alignment of Multiple Texts

At the simplest conception, multiple granularities of text are already supported as each word could be associated with the Canvas via a dedicated Annotation, each line could have its own Annotation, and so on, multiplied by as many transcriptions, editions and translations as are available. However, a user interface would benefit from guidance as to which Annotations should be displayed together, especially if the text is to be presented in a different window or frame from the image that depicts it, and is derived from multiple sources via multiple annotations.

In the case of multiple diplomatic transcriptions, the Annotations that contain the various transcriptions of a given word or line have a common segment of the Canvas as their Target. Hence, an existing SharedCanvas construction can be used. In this scenario a single Annotation is used with a single Target of the word or line, and the Body consists of a TextChoice with one option per transcription.

However, if the texts are a transcription, an edition and a translation, the segmentation of the Canvas will not be identical. In particular, a translation will be at the sentence or paragraph level to ensure readability, even if an edition only makes changes at the individual word level. As there is different segmentation, multiple Annotations will somehow require alignment.

To solve this use case, it is necessary to introduce a new type of resource at the discovery level in order to explicitly group together the annotations for each text. This new construction is called a Layer; each complete text would have its own Layer, and have the same ordered Aggregation structure as the other discovery lists. The Layer would aggregate either Annotations directly or the existing Lists of Annotations. Thus, the diplomatic transcription annotations would be aggregated in one Layer, the edition in a second, and a hypothetical translation in a third. The Layer construction could also be used for images, audio files or any other collection of resources that should be displayed together. These Layers are depicted as the 'ImgLyr', 'Txt1Lyr' and 'Txt2Lyr' nodes in Figure 7.

The alignment between texts to implement highlighting is most easily solved by calculating the overlap between the annotation segments on the canvas. When the user clicks on a line of text in the transcription, the system would first find and highlight the annotated region on the canvas, then look for other annotations that cover the same area and highlight the equivalent sections of their texts.

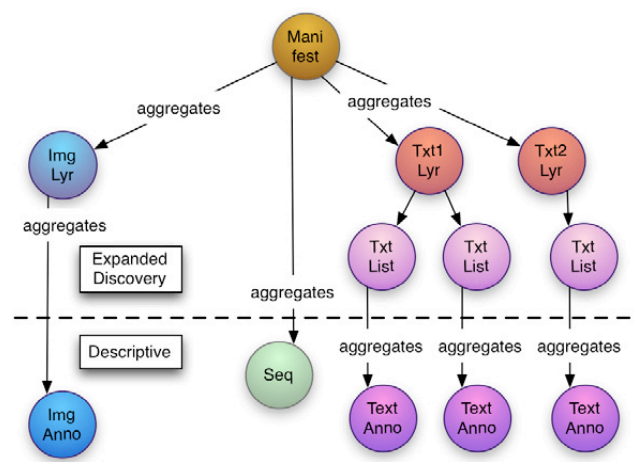

Figure 7. Expanded Discovery with Layers

**Interaction with Canvas Segments**
The final requirement specific to SharedCanvas is to model the complex layouts of the Sailing Letters in a way that allows a viewer to interact with the material in a rich and meaningful way. The two interactions described both involve manipulation of a segment of a Canvas, either to rotate it to the correct reading orientation, or to hide and show it as a means of simulating the folding of the letter.

The Zone construction from the SharedCanvas model provides the solution for these requirements. A Zone creates a new space that can be annotated onto one or more Canvases and groups other annotations along with their relative layout. A Zone that covers the segment that needs to be rotated could be created, and all of the Annotations for that segment would then target the Zone rather than the Canvas. The user interface could expose a rotate option, and rotate the rendering of the Annotations around an axis of the Zone, given the number of degrees.

In Figure 8, the text in the bottom right hand corner needs to be rotated clockwise by 270 degrees to be at the correct reading angle, however the text in the upper half is already in the normal orientation. The Zone collects the TextAnnotations and an ImageAnnotation for part of the image, and associates it with the equivalent part of the Canvas. The angle is recorded in the Zone with the readingAngle property, for use by the user interface.

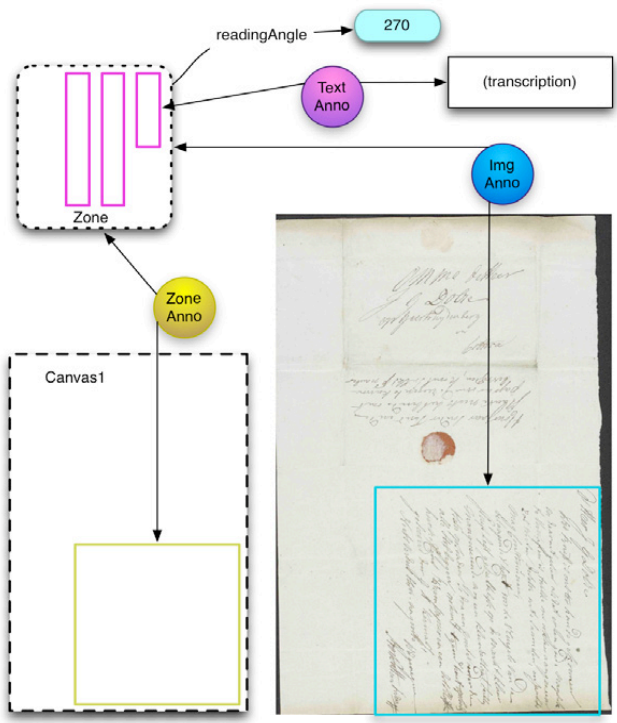

Figure 8. Rotation of Zone of Canvas

For folding and unfolding of the letters, a ZoneChoice would be used, as presented in Figure 9. Either the 'Zone-Folded' Zone would be rendered, with no image or text as this represents the area folded back underneath the rest of the letter and hence out of sight, or the Zone-Unfolded Zone would be used with the image and text being displayed. In Figure 9, only the ImageAnnotation is depicted as 'ImgAnno', which associates the section of image with the Zone. TextAnnotations would also be present in the full model to transcribe the address, in this case. The user interface would have to present this choice of Zones to the user, potentially with several other such dynamic options for other folds. If further digitizations of the letter were to be made in the future, then images of the folded up states might be available, rather than simply displaying an empty Zone.

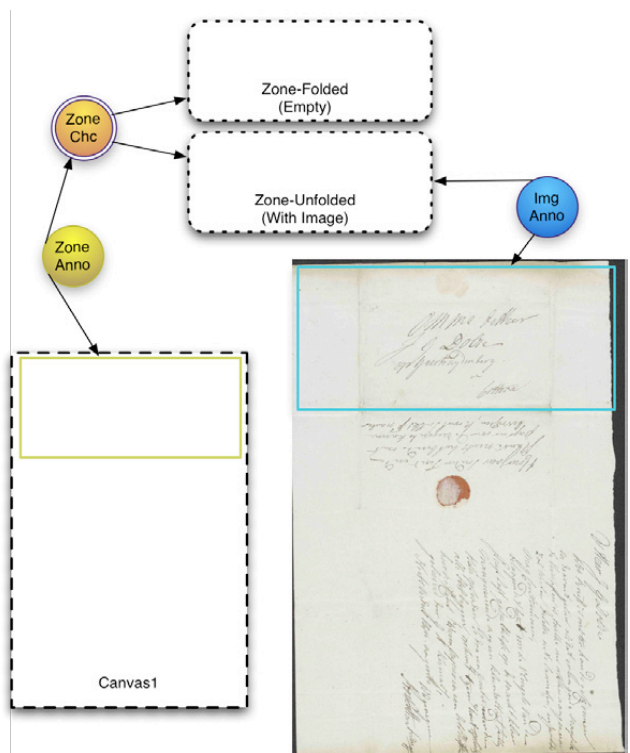

Figure 9. ZoneChoice for Dynamic Interaction

**Open Annotation Requirements**
The two use cases that remain are not directly tied to the SharedCanvas model, but instead rely on features of the underlying Open Annotation model. Annotations on the resources painted on to Canvases are not considered as part of the scope of SharedCanvas, as any resource can be annotated in this way. The use cases are:
1. Annotation of Textual Resources
2. Reuse of Segment Descriptions

**Annotation of Textual Resources**
The ability to annotate texts with semantic information is part of the Open Annotation guidelines. By creating ConstrainedTargets that identify a section of the text, and Constraints that describe where that section is located, it is possible to then associate further information with the identifier for that section (the ConstrainedTarget). For semantic annotations, where the information is intended for machine consumption, the Open Annotation model reuses the identity of the section within an RDF statement in the Body of an Annotation. This is an extensible and explicit method that does not require creating new relationships from the Annotation, and at the same time allows the association of metadata, such as the creator and time of creation or publication, with the statement itself and separate from the creation and publication time of the Annotation.

This covers use cases of Named Entity Recognition, both automated via machine learning techniques and manual, as well as modeling of the text rather than the Canvases into different hierarchies. The section of text containing the name of a person or other entity becomes the Target of an Annotation. The Body of that Annotation is an RDF statement that relates the section's identifier with the identifier of the entity. This is depicted in Figure 10 in which the RDF statement says that ct2 (the section's identifier) references someone called 'Carel'. Similarly, the section's identifier could be associated with a named hierarchical text structure as is the case with the example of a Sentence in the annotation on the ct1 segment in Figure 10. The user interface could then take action based on this information, perhaps by linking all mentions of the entity together, or allowing the user to navigate through the text, and hence the Canvases to which the texts are annotated, via the hierarchy described.

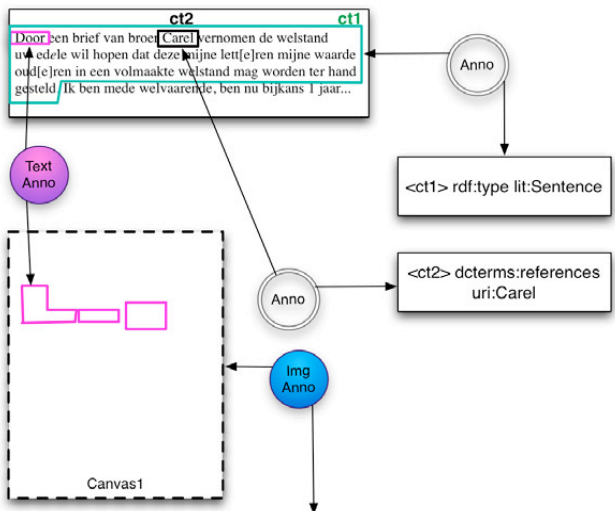
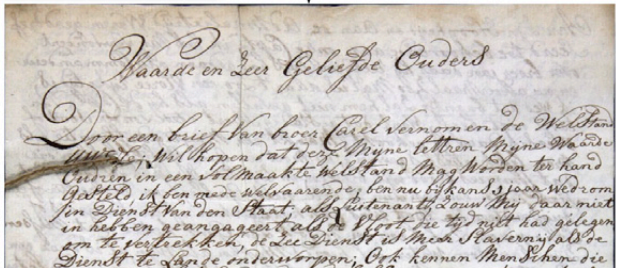

Figure 10. Semantic Annotations of Text

**Reuse of Segment Descriptions**

The final requirement is about reusing segmentation information in a template style manner, rather than recreating new segments for every new Canvas. This is supported by the Open Annotation model, due to the distinction between Constraints that describe an area in a reusable manner, and ConstrainedTargets and ConstrainedBodies that identify that segment within a particular resource.

In order to describe the top left hand cell in one of the KdK pages, as per Figure 11, only the x and y coordinates of the top left hand corner, and the height and width of the area are required. This could be expressed either in absolute pixel counts, but expressing it in percentages of the total height and width of the image is more likely to be easily reusable across images or Canvases of different dimensions. The information would be recorded in a Constraint, depicted as node 'C' in Figure 11, with its own URI that would be then be referenced from multiple Annotations, such as TextAnn1 and TextAnn2, and applied to different Canvases, page 348 and 350 in this case. Using this method, the layout could be described once without reference to a specific image or Canvas, and then reused for all of the digitized pages that follow that particular layout.

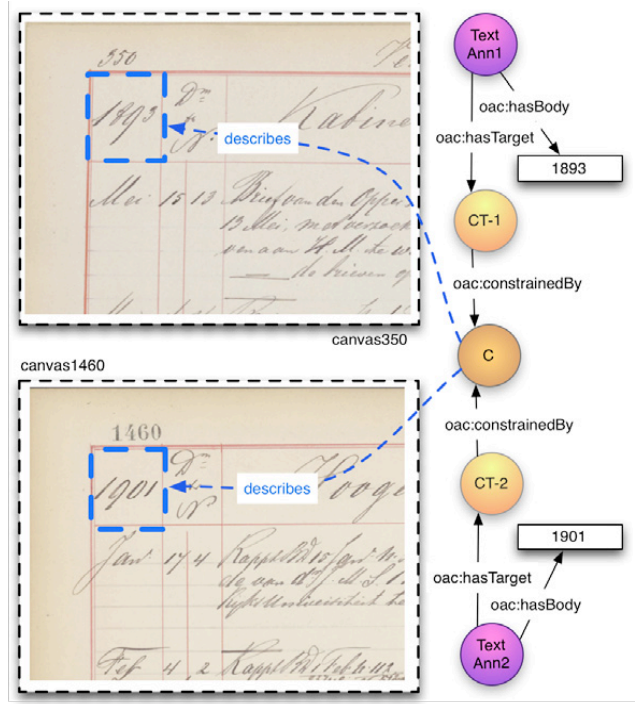

Figure 11. Reuse of Segment Descriptions

## Conclusions

The SharedCanvas data model has at its heart the complex relationships between texts, the digital images of textual content and the often unique physical objects, such as medieval manuscripts, that have preserved the texts from ages past. Its goal is to provide a single input to digital facsimile systems such as page-turners, in order to provide cross-repository interoperability in a distributed environment.

In this paper we first described the model and the underlying annotation and aggregation ontologies on which it is built, and two previously unknown collections. The use cases of the collections were then applied in order to evaluate the model's general applicability.

It was possible to describe all of the use cases using the model, despite their complexity. Three were handled at the level of the SharedCanvas model itself, whereas two others were more appropriately tackled at the level of the underlying Open Annotation model. One new construction was introduced at the discovery level: a Layer that groups annotations that semantically belong together. This is a refinement of the existing discovery model, which takes user interface requirements into account and is fully backwards compatible with existing uses.

This result is particularly important as a validation of the approaches taken by the SharedCanvas and Open Annotation models. As it was possible to successfully describe these previously unknown collections and requirements, the likelihood is high that they will be applicable to many other scenarios without further intervention or significant extension.

---

[i] http://www.kabinetderkoningin.nl/nl/kdk111.html
[ii] http://www.kb.nl/sl/index.html